\documentclass[12pt,preprint]{aastex}

\begin{document}

\title{Spatially Resolved Spectroscopic Observations of a Possible E+A
Progenitor SDSS J160241.00+521426.9}

\author{K. Matsubayashi\altaffilmark{1}, M. Yagi\altaffilmark{2},
T. Goto\altaffilmark{3, 4}, A. Akita\altaffilmark{1}, H. Sugai\altaffilmark{1},
A. Kawai\altaffilmark{1}, A. Shimono\altaffilmark{1},
T. Hattori\altaffilmark{4}}
\altaffiltext{1}{Department of Astronomy, Kyoto University, Kitashirakawa
Oiwakecho, Sakyo-ku, Kyoto 606-8502, Japan; kazuya@kusastro.kyoto-u.ac.jp}
\altaffiltext{2}{National Astronomical Observatory of Japan, 2-21-1 Osawa,
Mitaka, Tokyo 181-8588, Japan}
\altaffiltext{3}{Institute of Astronomy, University of Hawaii, 2680 Woodlawn
Drive, Honolulu, Hawaii 96822-1897, USA}
\altaffiltext{4}{Subaru Telescope, National Astronomical Observatory of Japan,
650 North A'ohoku Place, HI, 96720, USA}

\begin{abstract}

In order to investigate the evolution of E+A galaxies, we observed a
galaxy SDSS J160241.00+521426.9, a possible E+A progenitor which shows both
emission and strong Balmer absorptions, and its neighbor galaxy.
We used the integral field spectroscopic mode of the Kyoto Tridimensional
Spectrograph (Kyoto3DII), mounted on the University of Hawaii 88-inch telescope
located on Mauna Kea, and the slit-spectroscopic mode of the Faint Object
Camera and Spectrograph (FOCAS) on the Subaru Telescope.
We found a strong Balmer absorption region in the center of the galaxy and an
emission-line region located 2 kpc from the center, in the direction of its
neighbor galaxy.
The recession velocities of the galaxy and its neighbor galaxy differ only by
100 km s$^{-1}$, which suggests that they are a physical pair and would have
been interacting.
Comparing observed Lick indices of Balmer lines and color indices with those
predicted from stellar population synthesis models, we find that a suddenly
quenched star-formation scenario is plausible for the star-formation history of
the central region.
We consider that star formation started in the galaxy due to galaxy
interactions and was quenched in the central region, whereas star formation in
a region offset from the center still continues or has begun recently.
This work is the first study of a possible E+A progenitor using spatially
resolved spectroscopy.

\end{abstract}


\section{INTRODUCTION}

E+A galaxies are understood as post-starburst galaxies due to the presence of
strong Balmer absorption lines (H$\beta$, H$\gamma$, H$\delta$) and the lack of
emission lines \citep{Poggianti:1999, Goto:2005}.
Examinations of galaxy morphologies \citep{Yang:2004, Yang:2008} and companion
galaxies statistics \citep{Goto:2005, Yamauchi:2008} led to the conclusion that
local field E+A galaxies are mainly driven by mergers and interactions.
Integral field spectrograph (IFS) observations for E+A galaxies indicate
disturbed morphologies and significant rotation, which supports the theory that
they are produced by gas-rich galaxy mergers and interactions
\citep{Pracy:2009}.
Recently, \citet{Rich:2010} has reported that a starburst galaxy NGC 839 has
both A-type stellar population and a galactic wind, and has discussed a link
between lower-mass starburst systems and E+A galaxies.
An E+A galaxy is one phase in galaxy evolution, and thus, it is important
to understand the evolution of E+A galaxies.

All these results, however, are based on information obtained from galaxies
already in the $post$-starburst phase.
In order to investigate the evolution of E+A galaxies, it is necessary to
examine pre-E+A galaxies, or E+A progenitors, which would have both starburst
and post-starburst regions, and this requires
spatially resolved spectroscopic observations.

In this work, we observed a possible progenitor of an E+A galaxy, SDSS
J160241.00+521426.9 (J1602), using the IFS mode of the Kyoto Tridimensional
Spectrograph \citep[Kyoto3DII:][]{Sugai:2010}, mounted on the University of
Hawaii 88-inch telescope on Mauna Kea.
The galaxy has an apparent companion galaxy, seen in the Sloan Digital Sky
Survey (SDSS) image.
The 3$\arcsec$-fiber spectra from the SDSS show an interesting combination of
very strong Balmer absorption lines and emission lines.
With the IFS observation on the UH 88-inch telescope as well as 
slit-spectroscopic observations using the Subaru Telescope Faint Object Camera
and Spectrograph \citep[FOCAS: ][]{Kashikawa:2002}, we investigated the
spatial distributions of absorption and emission line regions.
Identifying the locations of starburst regions in E+A galaxies provides
constrains on E+A galaxy formation theories.
The presence of both post-starburst regions and current starburst regions in
one galaxy may provide us with an unprecedented opportunity to examine the
progenitor of an E+A galaxy.
Through detailed analysis, we attempted to understand the underlying physical
process that caused the simultaneous presence of post-starburst and current
starburst regions in this galaxy.

\section{OBSERVATIONS AND DATA REDUCTION}
\label{sec:obs}

\subsection{Target}
\label{sec:obs-target}

We selected a target galaxy which has both strong post-starburst
and starburst properties, based on its line ratios from an H$\delta$ strong
galaxies catalog \citep{Goto:2005} produced from the spectra of the Sloan
Digital Sky Survey Data Release 2 \citep[SDSS DR2:][]{Abazajian:2004}.
The redshift of J1602 was measured as 0.0430 $\pm$ 0.0001 \citep{York:2000},
and the distance was estimated as 170 Mpc (1\arcsec = 0.82 kpc) with h = 0.73,
$\Omega_m$ = 0.24, and $\Omega_{\Lambda}$ = 0.76 \citep{Spergel:2007}.
A possible companion galaxy appeared in the same field of view; its redshift
had not been measured by the SDSS.
We retrieved a K$_s$-band fits image from 2MASS \citep{Skrutskie:2006} and
estimated the total K$_s$-band magnitudes of J1602 and the companion galaxy as
14.8 mag and 15.7 mag, respectively.
Using mass-to-luminosity ratio of 0.6 \citep{Balogh:2005}, we calculated the
stellar masses of J1602 and the companion galaxy as 3.4 $\times$ 10$^9$
M$_{\odot}$ and 1.5 $\times$ 10$^9$ M$_{\odot}$, respectively, assuming that the
redshift of the companion galaxy is the same as that of J1602.

\subsection{Kyoto3DII Data}

We observed J1602 on 2005 May 25 (UT) using the IFS mode of the Kyoto3DII
mounted on the University of Hawaii 88-inch (2.2m) telescope at Mauna Kea.
This mode uses a 37$\times$37 lenslet array, allowing us to obtain
simultaneous spectra of $\sim 10^3$ spatial elements.
The Kyoto3DII has two separate fields of view (FOV): one for the target field
and another, smaller FOV for the sky field.
This is important for accurate sky subtraction.
The FOV of the array for the target field is $\sim 16\arcsec \times 12\arcsec$,
with a spatial sampling of 0$\arcsec$.43.
We used the No.2 grism and No.2 filter, which cover the 4170 -- 5260 \AA\,
wavelength range \citep{Sugai:2010}, and obtained two 1-hr exposure frames. 
The wavelength resolution in a full-width half maximum was $\sim$ 4.8 \AA,
which corresponds to a velocity resolution of $\sim$ 290 km s$^{-1}$.

The IFS data of a spectroscopic standard star from IRAF's irscal database,
HD161817, were used for flux calibration, and those of helium and halogen lamps
were used for wavelength calibration and flat-fielding, respectively.
The velocity accuracy was estimated to be 22 km s$^{-1}$ (1$\sigma$) from the
lines of the comparison lamp and sky absorption.
The averaged sky spectrum was subtracted from the target spectra.
Cosmic rays were removed.
Atmospheric dispersion correction was performed for all object frames.
The spatial resolution was $\sim$ 1$\arcsec$.8, which was measured from the
FWHM of a standard star.
The background noise level was 1.0 $\times$ 10$^{-18}$ erg cm$^{-2}$ s$^{-1}$
\AA$^{-1}$ (0.43 arcsec)$^{-2}$ (1$\sigma$). 
We tried Voronoi binning \citep{Cappellari:2003} for Kyoto3DII data, but quick
drop off of continuum flux in outer region led to very large bins, and thus
we did not use Voronoi binning for Kyoto3DII data.

\subsection{FOCAS Data}

We also observed J1602 and the neighbor galaxy on 2009 May 20 (UT) using FOCAS
on the Subaru Telescope.
We adopted a longslit of 0\arcsec.8 width, and observed with the 300B grism
without a filter.
The wavelength coverage was about 3750 \AA\, -- 7100 \AA\, and the spectral
resolving power was $\sim$ 700.
The CCD was read out in 3 $\times$ 1 binning, and the image scale along the
slit was 0.31 arcsec pixel$^{-1}$.
We obtained three sets of 10 min exposures along the major axis of J1602, and
another three sets of 5 min exposures along the major axis of the neighbor.

We used five dome-flat frames for flat fielding.
A spectrophotometric standard star from \citet{Oke:1990}, BD+28D4211, was
observed for flux calibration.  
The seeing size, estimated from the standard star, was 0\arcsec.56.  
The wavelength of each exposure was calibrated using a ThAr lamp. 
The velocity accuracy was estimated to be 17 km s$^{-1}$ (1$\sigma$) from
the lines of the comparison lamp and sky emission.
The averaged sky spectrum was subtracted from target spectra.
For investigating the spatial variations of Lick indices, we combined adjacent
spatial pixels until the S/N ratios of combined pixels reached the target S/N
(cf. Voronoi binning: \citealt{Cappellari:2003}).
Target S/N ratios were set as 100 for J1602 and as 50 for the companion galaxy.

\subsection{Stellar Population Synthesis Models}
\label{sec:BC03-model}

In order to predict star-formation history in several regions in the galaxies,
we compare in Section \ref{sec:discuss-age} the observed Lick indices of the
H$\gamma_A$ and H$\delta_A$ \citep{Worthey:1997} and the color indices in the
rest-frame with model values. 
We used stellar population synthesis models \citep[GALAXEV:][]{Bruzual:2003}.
Each model provides the Lick indices and the color indices as a function of
starburst age, assuming a star-formation history, the Salpeter IMF 
\citep{Salpeter:1955}, and no dust.
We calculated three models as follows:
(1) a simple stellar population (SSP) model, in which a starburst occurred only
at 0 yr,
(2) an exponential (exp) model, in which the star-formation rate decreased
exponentially (on a time scale $\tau$ = 1 Gyr), and
(3) a constant (const) model, in which starbursts continued constantly.
Color indices are calculated from redshift-corrected model spectra.
Model spectra do not include emission lines so that emission line correction is
needed for the observed color indices in starburst regions.

\subsection{SDSS Data}

In order to calculate color indices, we retrieved the SDSS DR7 corrected images
\citep[FITS files:][]{Abazajian:2009} and performed aperture photometry using a
1\arcsec.7-diameter aperture at the several regions in the galaxies.
The FITS images were re-sampled using SWarp \citep{Bertin:2002} according to
the WCS information of each corrected image.
SWarp also subtracted the background.
The zero point for the photometry was calibrated with tsField\footnote{
http://www.sdss.org/dr7/algorithms/fluxcal.html}.
Galactic extinction was corrected according to \citet{Schlegel:1998}, adopting
E(B-V) = 0.015 for J1602.

As we will compare the color indices with dust-free and emission-free models by
\citet{Bruzual:2003}, the internal extinction and the emission line corrections
were performed where emission lines were detected.
For internal extinction corrections, we used the value estimated from the
H$\alpha$/H$\beta$ flux ratio observed by FOCAS because the H$\gamma$ flux
observed with Kyoto3DII was affected by absorption and a low signal-to-noise
ratio.
The H$\alpha$/H$\beta$ flux ratio was calculated from the Galactic extinction
corrected values.
The extinction law of \citet{Calzetti:2000} was used for extinction correction
for J1602 and its companion galaxy.
We used two reddening ratios between gas and stars:
$E_{star}(B-V)$ = $E_{gas}(B-V)$, which corresponds to a situation that dust
exist outside of a star-forming region, and
$E_{star}(B-V)$ = 0.44 $E_{gas}(B-V)$ \citep{Calzetti:2000}, which corresponds
to a situation that dust coexist with stars and gas in a star-forming region.
For the emission line correction, the spectra observed with FOCAS were used.
We calculated the fraction of the emission line fluxes from measured equivalent
widths, and corrected the magnitudes in each band.
This correction was performed for emission lines within the FOCAS wavelength
range of 3750 -- 7100 \AA\, only.
Thus, the corrected u-g indices may be redder, and r-i may be bluer.
As we cannot correct the effect of emission lines in the z-band, no emission
line correction for the i-z indices was attempted.

\subsection{Lick Indices}

The Lick indices were calculated from Kyoto3DII and FOCAS spectra after their
spectral resolution was matched to the Lick system using Fig. 7 and Table 8 of
\citet{Worthey:1997}.
Lick offsets displayed in Table 6 of \citet{Bruzual:2003} (Table
\ref{tb:Lick-offsets}) were applied to the observed values, because only one and
no Lick standard star was observed by Kyoto3DII and FOCAS, respectively, in our
observing runs.
The Lick offset of H$\gamma_A$ with Kyoto3DII was estimated from HD161817 data
as -0.20 $\pm$ 0.02 \AA.
Errors of Lick indices were estimated using equations of \citet{Cardiel:1998}.

\section{RESULTS}
\label{sec:result}

\subsection{Kyoto3DII Results}

\subsubsection{Region Definition}
\label{sec:result-image-3dii}

\begin{figure}
\epsscale{.90}
\plotone{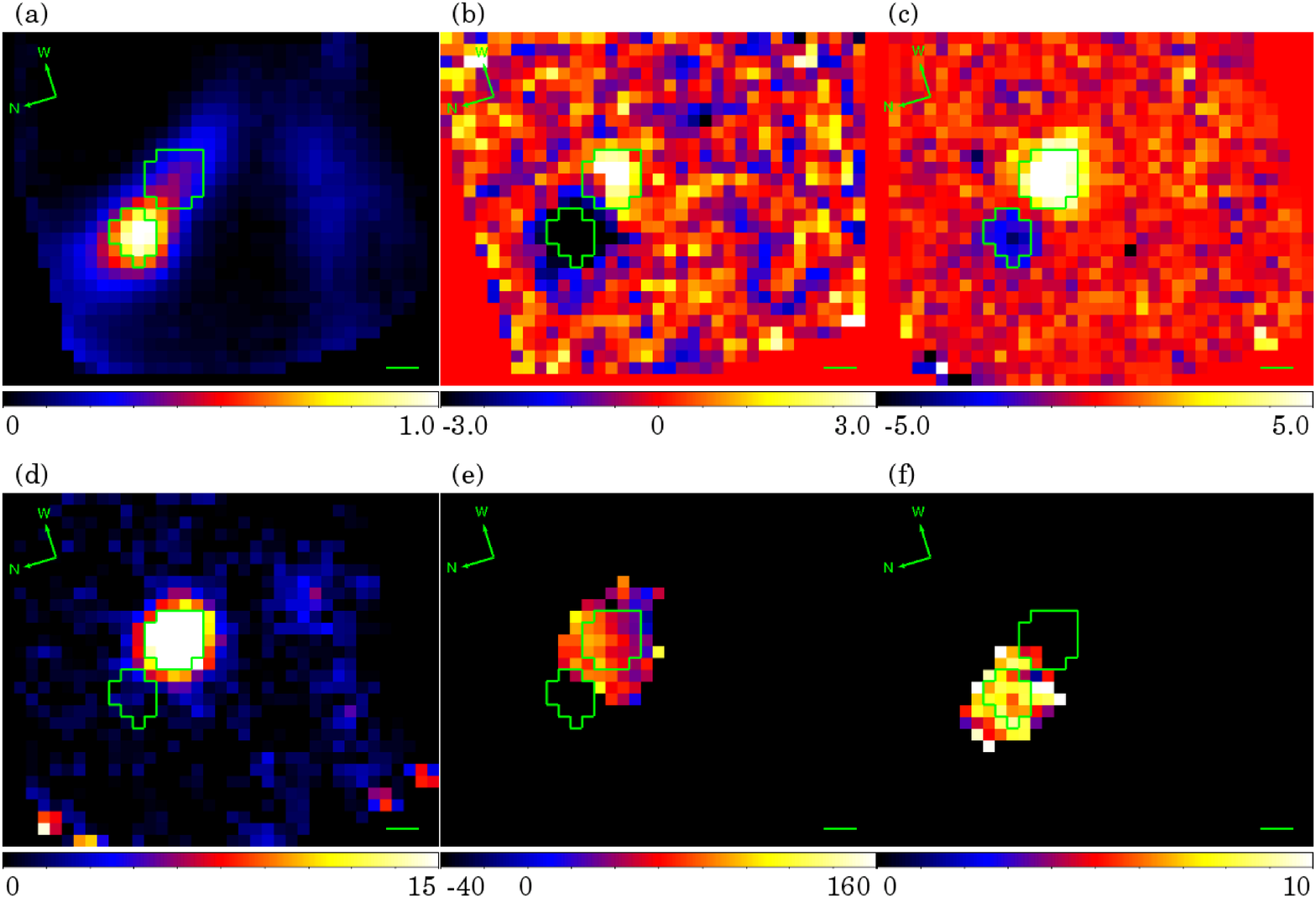}
\caption{(a) Continuum flux (4600 -- 5000 \AA\, in observed frame), (b)
H$\gamma$ flux, (c) H$\beta$ flux, (d) [\ion{O}{3}]$\lambda$5007 flux, (e)
[\ion{O}{3}]$\lambda$5007 velocity with respect to the systemic velocity 
(z = 0.043), and (f) H$\gamma_A$ Lick index maps of SDSS J160241.00+521426.9
(J1602) and the companion galaxy observed with Kyoto3DII.
The unit of flux density in the continuum image is 10$^{-17}$ erg cm$^{-2}$
s$^{-1}$ \AA$^{-1}$ (0.43 arcsec)$^{-2}$, and that of flux in the H$\gamma$,
H$\beta$, and [\ion{O}{3}]$\lambda$5007 images is 10$^{-17}$ erg cm$^{-2}$
s$^{-1}$ (0.43 arcsec)$^{-2}$.
The unit of [\ion{O}{3}]$\lambda$5007 velocity map is km s$^{-1}$, and that of
H$\gamma_A$ Lick index map is \AA.
The two arrows in the upper left indicate the north and west directions.
The bar at the lower right in each panel indicates a 1 kpc length.
The two regions surrounded by the lines represent PS1 (left) and SB1 (right) in
the Kyoto3DII definition.
(A color version of this figure is available in the online journal.)
\label{fig:image-3dii}}
\end{figure}

\begin{figure}
\epsscale{.80}
\plotone{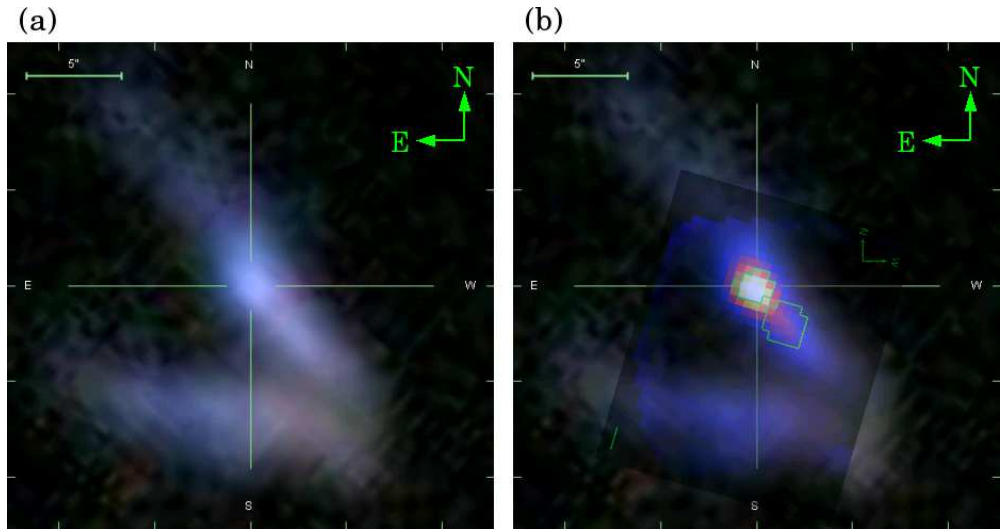}
\caption{(a) Original SDSS image.
North is up and east is left.
The bar in the upper left indicates 5\arcsec\ in length.
(b) Continuum image observed with Kyoto3DII superposed on SDSS image.
(A color version of this figure is available in the online journal.)
\label{fig:3dii-sdss}}
\end{figure}

Figure \ref{fig:image-3dii}(a) shows the continuum image in the 4600 \AA\, --
5000 \AA\, wavelength range in observed frame obtained from Kyoto3DII IFS data.
Two galaxies fall within our FOV: a brighter northern galaxy and a fainter
southern one.
We designated the northern galaxy J1602 and the southern one, the possible
companion galaxy of J1602.
The positions of the galaxies in our continuum image are consistent with those
in the image retrieved from the SDSS archive data (Fig. \ref{fig:3dii-sdss}).
As described below in detail, absorption lines were found in some regions
(e.g., the H$\gamma$ and H$\beta$ maps: Figs. \ref{fig:image-3dii}(b) and (c),
respectively), and emission lines were found in other regions in both galaxies
(e.g., the [\ion{O}{3}]$\lambda$5007 map: Fig. \ref{fig:image-3dii}(d)).
From the observed wavelengths of the H$\beta$, H$\gamma$, and
[\ion{O}{3}]$\lambda\lambda$4959,5007 emission lines, the redshift of J1602
was measured as 0.0431 $\pm$ 0.0001 (1$\sigma$), consistent with the results
from the SDSS spectroscopic data.
We also calculated the redshift of the fainter galaxy from its
[\ion{O}{3}]$\lambda\lambda$4959,5007 emission lines as 0.0434 $\pm$ 0.0002.
The redshift of the possible companion galaxy is almost the same as that of
J1602, showing that these two galaxies are indeed physically close to each
other and, most likely, form a dynamically interacting pair.
As this companion galaxy was not spectroscopically observed with the SDSS,
our spectroscopic results confirmed for the first time a physical connection
between these galaxies.

We defined two regions in J1602 (Fig. \ref{fig:image-3dii}): the Balmer
absorption region (post-starburst region; hereafter PS1) and the [\ion{O}{3}]
emission region (starburst region; hereafter SB1).
These regions were defined by visual inspection of the 3-dimensional data.
The location of PS1 lies at and around the center of J1602, whereas SB1 is
located 2\arcsec.5 southwest from PS1.
The apparent sizes of PS1 and SB1 are 2.6 arcsec$^2$ and 4.1 arcsec$^2$,
respectively.
The size of PS1 is comparable to the seeing size ($\sim$ 2.5 arcsec$^2$),
whereas that of SB1 is significantly larger than the PSF.
Tables \ref{tb:3dii-reg1} and \ref{tb:3dii-reg2} provide detailed data for each
absorption or emission line measured by Gaussian fitting in PS1 and SB1,
respectively.

\subsubsection{PS1}
\label{sec:result-reg1-3dii}

\begin{figure}
\epsscale{1.00}
\plotone{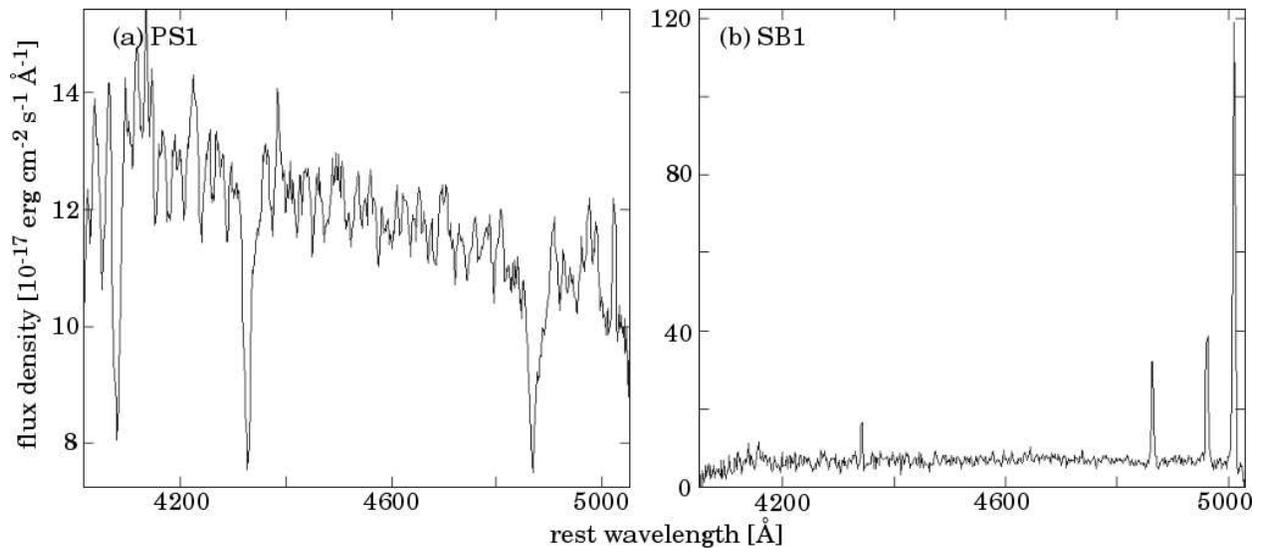}
\caption{(a) The smoothed spectrum at PS1 observed with Kyoto3DII.
(b) The SB1 spectrum observed with Kyoto3DII.
\label{fig:spec-3dii}}
\end{figure}

Figure \ref{fig:spec-3dii}(a) shows the spectrum of PS1.
H$\beta$, H$\gamma$, and H$\delta$ lines are seen in absorption.
No emission line was found.
Strong Balmer absorption lines (Lick(H$\gamma_A$) = 8.3 \AA\,: Table
\ref{tb:FOCAS-Lick}) and no emission line suggest that PS1 is a post-starburst
region.
We will discuss the age and star-formation history of this region in Section
\ref{sec:discuss-age}.
Figure \ref{fig:image-3dii}(f) shows that H$\gamma_A$ Lick index outside PS1 is
also larger than $\sim$ 5 \AA, which suggest that post-starburst population
presents there.

\subsubsection{SB1}
\label{sec:result-r2-3dii}

The spectrum of SB1 is quite different from that of PS1.
Figure \ref{fig:spec-3dii}(b) shows the SB1 spectrum.
The continuum emission in SB1 was fainter than that in PS1, and no absorption
lines were detected.
Instead, the [\ion{O}{3}]$\lambda\lambda$4959,5007, H$\beta$, and H$\gamma$
emission lines were detected, which indicates that SB1 is a starburst region.
The H$\gamma$ emission line was observed at a low S/N, probably affected by
stellar H$\gamma$ absorption (see Section \ref{sec:result-r2-focas}).
Although the [\ion{O}{3}]$\lambda$5007/H$\beta$ ratio in this region ($\sim$ 3)
is similar to typical AGN values, we doubt that there is any AGN activity in
this galaxy, because the [\ion{N}{2}]/H$\alpha$ and [\ion{S}{2}]/H$\alpha$
ratios from SDSS observation data are much smaller than theoretical 'maximum
starburst lines' defined by \citet{Kewley:2001}.
Moreover, given that the observed H$\alpha$ flux is reduced by stellar
absorption, the intrinsic [\ion{N}{2}]/H$\alpha$ and [\ion{S}{2}]/H$\alpha$
ratios are probably even smaller.

The star formation rate was estimated with two different assumptions.
From the H$\gamma$/H$\beta$ flux ratio, we estimated the extinction as 3.0 mag
in the V-band using the intrinsic H$\gamma$/H$\beta$ ratio of 0.466
\citep{Osterbrock:2006} and the \citet{Calzetti:2000} extinction law.
The star-formation rate in SB1, calculated from the extinction-corrected
H$\beta$ flux in 4.1 arcsec$^2$, which corresponds to 3.4 kpc$^2$, was
estimated as 2.4 M$_{\odot}$ yr$^{-1}$ \citep{Kennicutt:1998}.
Meanwhile, if we assume that the observed H$\beta$ and H$\gamma$ fluxes
represent the combination of emissions and absorptions whose equivalent
widths were the same as those of PS1, we can estimate the extinction as $\sim$
0 mag.
If we assume this extinction value, the star-formation rate would be 0.11
M$_{\odot}$ yr$^{-1}$.
In this estimate, we assumed the Salpeter initial mass function.

From our SB1 IFS data, we found spatial variation in the velocity centroid
derived from the [\ion{O}{3}]$\lambda$5007 emission line.
Figure \ref{fig:image-3dii}(e) shows [\ion{O}{3}]$\lambda$5007 velocity map
relative to redshift derived from SDSS spectrum.
The velocity gradient along the SSE-NNW direction was detected.
The measured velocities were -50 km s$^{-1}$ in the SSE and +30 km s$^{-1}$ in
the NNW with respect to the velocity center of SB1.
Similar velocity fields were detected in H$\beta$ and [\ion{O}{3}]$\lambda$4959,
although their fluxes were weaker than that of [\ion{O}{3}]$\lambda$5007.
This gas motion may be a local rotation or outflow/inflow.

\subsection{FOCAS Results}

\subsubsection{Region Definition}
\label{sec:result-image-focas}

\begin{figure}
\epsscale{.80}
\plotone{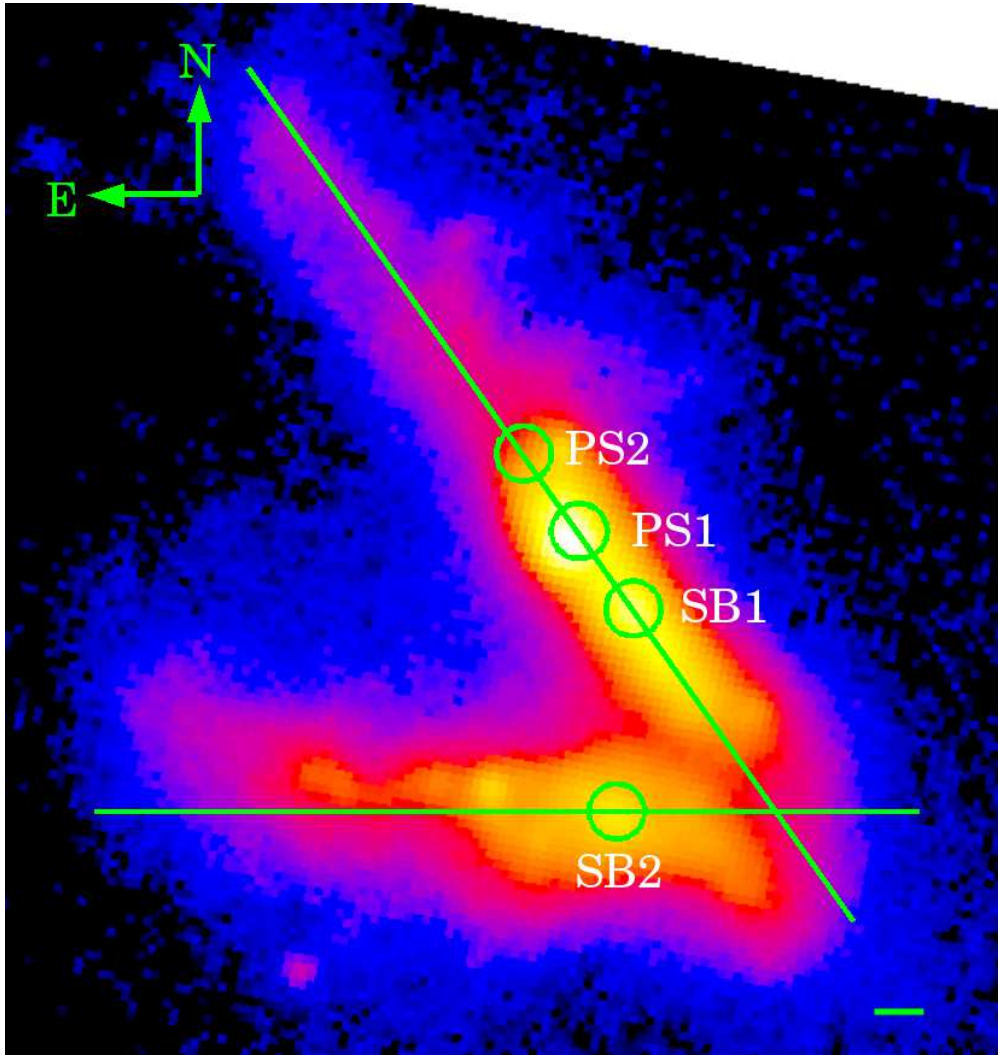}
\caption{Continuum image of J1602 observed with FOCAS presented on a
logarithmic scale.
North is up and east is left.
The lines over the galaxies display the slit positions.
Circles represent the position of each region.
The bar at the lower right represents a length of 1 kpc.
(A color version of this figure is available in the online journal.)
\label{fig:image-focas}}
\end{figure}

Figure \ref{fig:image-focas} shows the filterless image observed with FOCAS.
The spatial resolution of this image (0\arcsec.6) is the best of our images.
The southwest side of J1602 is brighter than the opposite side.
The difference in stellar mass or mass/luminosity ratio due to differences in
stellar ages may have produced this asymmetry.
The shape of the companion galaxy is also asymmetric.
The eastern side of this galaxy is 5 kpc longer than the western side, if we
consider that the continuum flux peak marks the galactic center.

\begin{figure}
\epsscale{.90}
\plotone{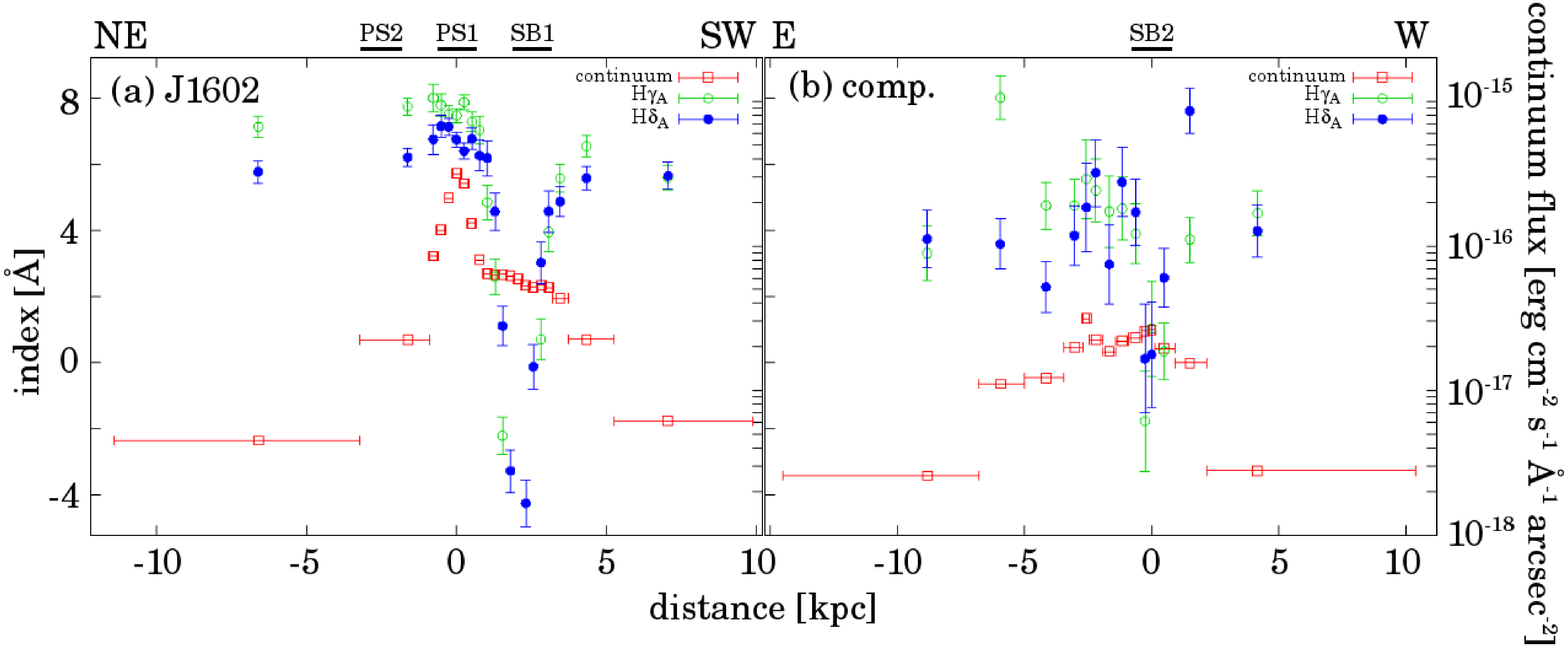}
\caption{Lick index profiles of (a) J1602 and (b) the companion galaxy observed
with FOCAS.
Filled and open circles represent H$\delta_A$ and H$\gamma_A$ Lick indices,
respectively, with 1 $\sigma$ error bars.
Open squares represent the continuum flux (4387 -- 4453 \AA\, in observed
frame) in logarithmic scale, as shown in the right axis.
The uncertainties for continuum flux are too small to be plotted in these
figures.
Horizontal bars represent binned regions.
The abscissa indicates distance from each galactic center, i.e., PS1
for J1602 and SB2 for the companion galaxy, respectively.
The locations of PS1, PS2, SB1, and SB2 are indicated on the top of the figure.
The characters on the top represent the directions relative to each galactic
center.
(A color version of this figure is available in the online journal.)
\label{fig:FOCAS-profile}}
\end{figure}

We defined three regions in J1602 in a manner similar to that used for the
Kyoto3DII definitions: the Balmer absorption region (PS1), the emission lines
region (SB1), and the region on the side opposite SB1 (PS2).
The location of PS1 is at the center of J1602.
The location of SB1 is 2\arcsec.5 southwest from PS1, and that of PS2 is
2\arcsec.5 northeast.
We also defined the central region in the companion galaxy as SB2.
To obtain one-dimensional spectra of these regions, we combined five pixels for
the spatial direction, which corresponds to 1\arcsec.6.
Tables \ref{tb:FOCAS-reg1}, \ref{tb:FOCAS-reg2}, \ref{tb:FOCAS-reg3}, and
\ref{tb:FOCAS-regA} provide detailed data for each absorption or emission line
measured by Gaussian fitting in PS1, SB1, PS2, and SB2, respectively.
Figure \ref{fig:FOCAS-profile} shows Lick index profiles of J1602 and the
companion galaxy from FOCAS data after binning low S/N pixels.

\subsubsection{PS1, PS2, and Outer Regions of J1602}
\label{sec:result-r1-focas}

\begin{figure}
\epsscale{.80}
\plotone{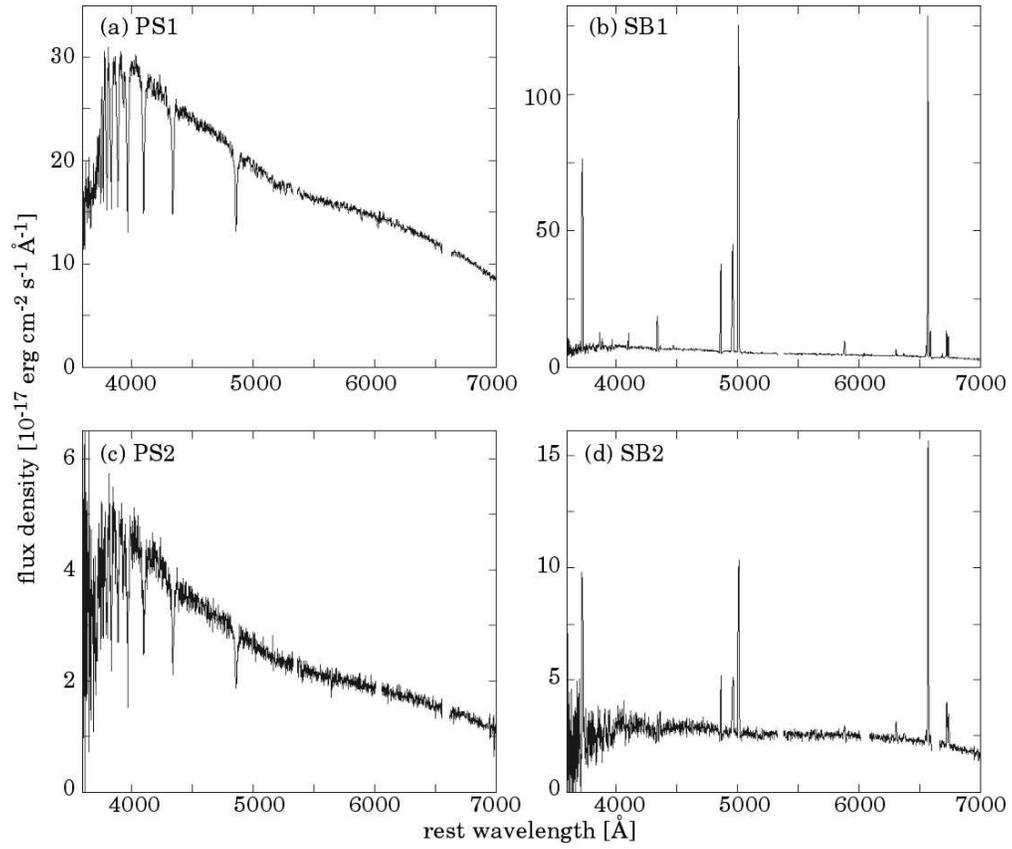}
\caption{Spectra at (a) PS1, (b) SB1, (c) PS2, and (d) SB2 observed with FOCAS.
\label{fig:spec-focas}}
\end{figure}

Figures \ref{fig:spec-focas}(a) and (c) display the spectra of PS1 and PS2,
respectively, observed with FOCAS.
We detected many absorption lines in PS1 and PS2 and found no emission line.
It is unclear whether the H$\alpha$ line is an absorption or emission line.
Table \ref{tb:FOCAS-Lick} shows H$\delta_A$ and H$\gamma_A$ indices of PS1 and
PS2.
Lick indices of H$\gamma_A$ and H$\delta_A$ are larger than 6 \AA\, in both
regions.
This suggests that not only PS1, but also PS2 is a post-starburst region.
We will discuss the star-formation history of PS1 and PS2 in detail in Section
\ref{sec:discuss-age}.

At outer regions of J1602, more than 5 kpc from the center, Lick indices of
H$\gamma_A$ and H$\delta_A$ are larger than 5 \AA\, (Fig. 
\ref{fig:FOCAS-profile}(a)).
These indices are similar to PS1 and PS2 values, and significantly larger than
those of old population with the age of $\sim$ 10 Gyr (see Section
\ref{sec:discuss-age}).
Thus, post-starburst population may be the main component on the continuum flux
around the H$\gamma$ wavelength at the whole J1602, including outer regions.
There are some examples which have post-starburst regions at the whole
galaxies, such as SDSS J161330.18+510335.5 \citep{Yagi:2006}.

\subsubsection{SB1}
\label{sec:result-r2-focas}

Figure \ref{fig:spec-focas}(b) shows the SB1 spectrum observed with FOCAS.
Many emission lines, such as the Balmer lines,
[\ion{O}{2}]$\lambda\lambda$3726,3729,
[\ion{O}{3}]$\lambda\lambda$4959,5007, [\ion{N}{2}]$\lambda\lambda$6548,6583,
and [\ion{S}{2}]$\lambda\lambda$ 6716,6731, were detected.
A closer inspection revealed that broad absorptions appear around the Balmer
emission lines, especially the H$\beta$.
Since the [\ion{N}{2}]$\lambda$6583/H$\alpha$ flux ratio was smaller than 0.1
(Table \ref{tb:FOCAS-reg2}), we concluded that no AGN activity was present in
SB1.

The extinction value was estimated from the H$\alpha$/H$\beta$ flux ratio,
although the H$\gamma$/H$\beta$ ratio was used in Section
\ref{sec:result-r2-3dii}.
This was done because the broad absorption around H$\gamma$ affects the
H$\gamma$ emission line flux measurement.
We estimated the extinction as 0.82 mag in V-band from the H$\alpha$/H$\beta$
ratio using the intrinsic H$\alpha$/H$\beta$ ratio of 2.87
\citep{Osterbrock:2006} and the \citet{Calzetti:2000} extinction law.
Using the H$\beta$ flux from the Kyoto3DII result and the extinction value from
the FOCAS result, the star-formation rate in SB1 was found to be 0.25
M$_{\odot}$ yr$^{-1}$.
Meanwhile, if we assume the same condition as Section \ref{sec:result-r2-3dii},
we can estimate the extinction as $\sim$ 0 mag and the star-formation rate as
0.11 M$_{\odot}$ yr$^{-1}$.

Then, we estimated the metallicity in this region from the
[\ion{N}{2}]$\lambda$6583/H$\alpha$ flux ratio, which was 0.077.
Using the relationship between the [\ion{N}{2}]/H$\alpha$ ratio and the
metallicity \citep{Nagao:2006}, the metallicity (12 + log(O/H)) was found to be
8.29 $\pm$ 0.09 (1$\sigma$).
This value is smaller than the solar value \citep[8.73:][]{Lodders:2009}, and
the metallicity Z was equal to 0.005.
This small metallicity may be explained by the fact that present stars have not
yet captured metals previous generation stars produced.
It might be the case that few stars, i.e., few metals have been produced.

\subsubsection{Companion Galaxy}
\label{sec:result-ra-focas}

Figure \ref{fig:spec-focas}(d) shows the SB2 spectrum observed with FOCAS,
which is similar to that of SB1.
Many emission lines, and the broad absorption around H$\beta$, were detected.
Because the [\ion{N}{2}]/H$\alpha$ ratio was smaller than 0.1 (Table
\ref{tb:FOCAS-regA}), we concluded that no AGN activity was present in this
galaxy.
Using the method outlined in Section \ref{sec:result-r2-focas}, we calculated
the extinction value of SB2 as 1.8 mag in the V-band from the
H$\alpha$/H$\beta$ flux ratio, and the metallicity (12 + log(O/H)) was
calculated as 8.11 $\pm$ 0.14 (1$\sigma$).
At the whole regions of the companion galaxy, Lick indices of H$\gamma_A$ and
H$\delta_A$ are $\sim$ 4 \AA\,, which are larger than old population (Fig.
\ref{fig:FOCAS-profile}(b)).
Thus, a post-starburst population may be present in some regions in the
companion galaxy.
Some additional emission line regions were found in this galaxy.
We conclude that these also indicate starburst regions because their line
ratios are similar to those of SB2.

\section{DISCUSSION}
\label{sec:discussion}

\subsection{Comparison of Models and Observed Values}
\label{sec:discuss-age}

We compared Lick indices of the H$\gamma_A$ and H$\delta_A$ and the color
indices in the rest-frame with predictions taken from star-formation models.
We used a metallicity of Z = 0.004 or 0.008, which is similar to the estimated
value from [\ion{N}{2}]6583/H$\alpha$ ratio (Section \ref{sec:result-r2-focas}).
Lick indices of H$\gamma_A$ and H$\delta_A$ are a good indicator of the
fraction of A-type stars.
\citet{LeBorgne:2006} displays that only the models with $\tau <$ 100 Myr pass
through the region with EW(H$\delta) >$ 6 \AA\, and
EW([\ion{O}{2}]$\lambda\lambda3726,3729) >$ -5 \AA\, (their Fig. 12).
The u-g, g-r, r-i, and i-z color indices generally increase with increasing
age.
The u-g color index is useful for determining the star-formation history in
each region because it has a strong dependence on star-formation history,
especially for ages of more than 30 Myr.

\subsubsection{PS1}

\begin{figure}
\epsscale{.80}
\plotone{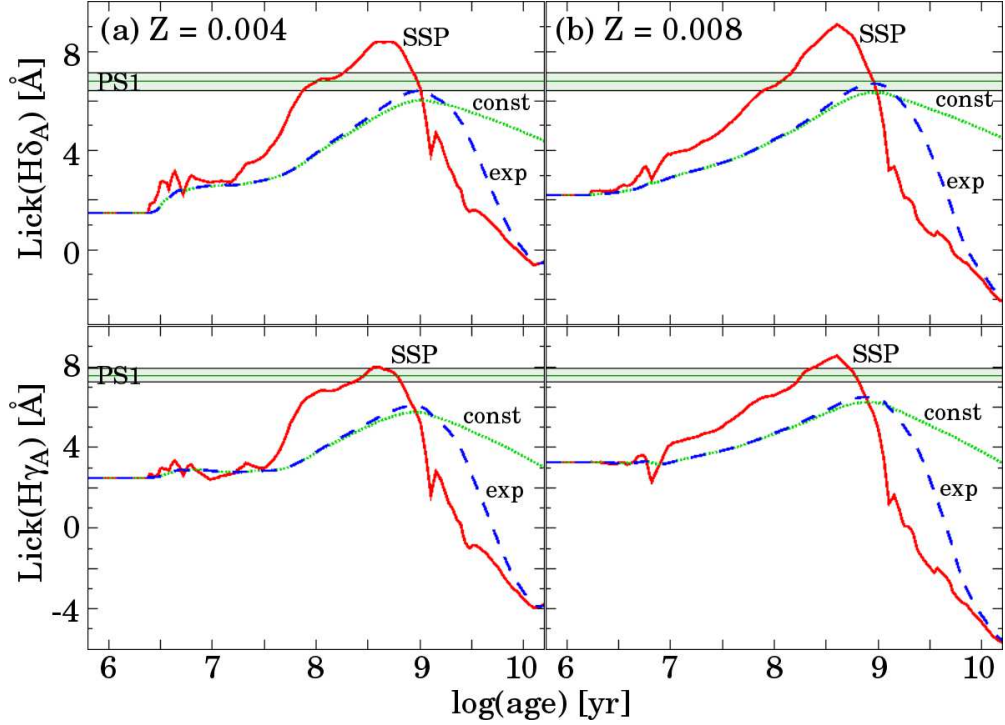}
\caption{H$\delta_A$ ({\it top}) and H$\gamma_A$ ({\it bottom}) Lick indices
predicted by three star-formation models \citep{Bruzual:2003} as a function
of starburst age, with the indices obtained in PS1.
The metallicities used in the models are (a) Z = 0.004 and (b) Z = 0.008,
respectively.
Solid, long-dashed, and dotted lines mark the Lick indices of SSP, exp, and
const models, respectively (see text for detail).
The faint horizontal lines mark the observed Lick indices in PS1 observed with 
FOCAS.
The shaded regions represent the values that fall within 3 $\sigma$ for FOCAS
data.
(A color version of this figure is available in the online journal.)
\label{fig:hgd-ew-model}}
\end{figure}

\begin{figure}
\epsscale{.80}
\plotone{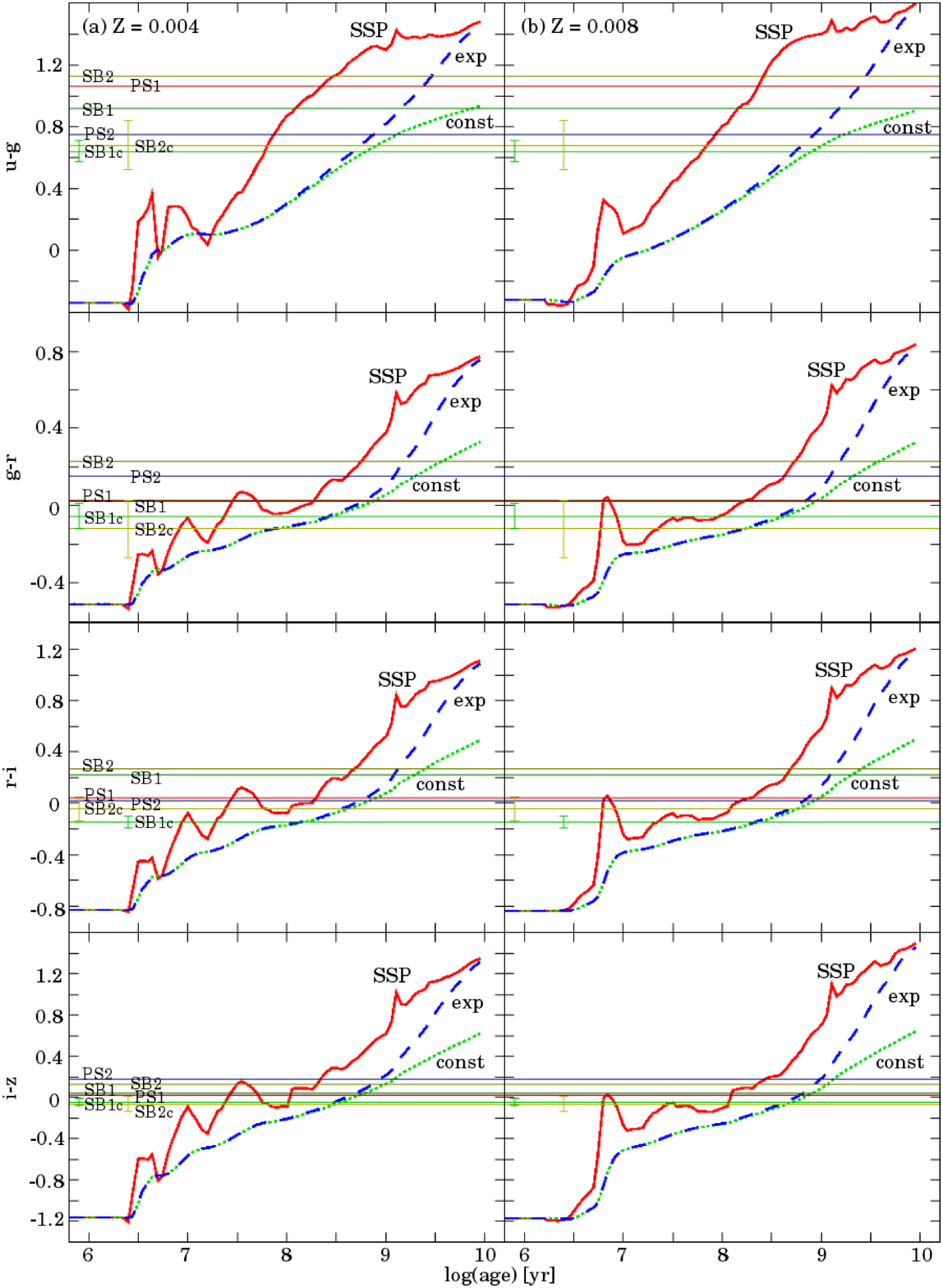}
\caption{Color indices predicted by three star-formation models
\citep{Bruzual:2003} as a function of starburst age, with the indices obtained
in PS1, PS2, SB1, and SB2.
The metallicities used in the models are (a) Z = 0.004 and (b) Z = 0.008,
respectively.
Solid, long-dashed, and dotted lines represent the color of the SSP, exp, and
const models, respectively (see text for detail).
Faint horizontal lines represent the observed and corrected indices.
Labels on these lines indicate the regions, e.g., the line marked "SB1"
represents the observed color index at SB1, while the one marked "SB1c"
represents the corrected value at SB1.
Bars on the SB1c and SB2c lines represent uncertainties due to the assumption
of gas-to-star reddening ratio.
(A color version of this figure is available in the online journal.)
\label{fig:color-model}}
\end{figure}

Figure \ref{fig:hgd-ew-model} shows the predicted Lick indices of H$\gamma_A$
and H$\delta_A$ as a function of starburst age.
The observed Lick indices in each region are summarized in Table 
\ref{tb:FOCAS-Lick} and are plotted in Fig. \ref{fig:hgd-ew-model}.
The large H$\gamma_A$ index in PS1 suggests that the SSP model at 100--700 Myr
is plausible; the const and exp model cannot produce these large values.
The H$\delta_A$ index of 6.8 \AA\, matches the SSP model at 0.1 or 1 Gyr, or
the const or exp models at 1 Gyr. 
Figure \ref{fig:color-model} shows the predicted u-g, g-r, r-i, and i-z
color indices as a function of starburst age.
The observed color indices in each region are summarized in Table
\ref{tb:color-sdss} and are plotted in Fig. \ref{fig:color-model}.
The observed u-g color in PS1, 1.06, corresponds to an age of 200 Myr in the
SSP model, to 2 Gyr in the exp model, and to longer than 10 Gyr in the const
model.
The other indices correspond to ages of $\sim$ 200 Myr in the SSP model or 700
Myr in the exp or const model.
Although an age of 7 Myr for Z = 0.004 or 30 Myr for Z = 0.008 in the SSP model
may fit the g-r, r-i, and i-z color indices, they completely disagree with the
red u-g color.

\begin{figure}
\epsscale{.90}
\plotone{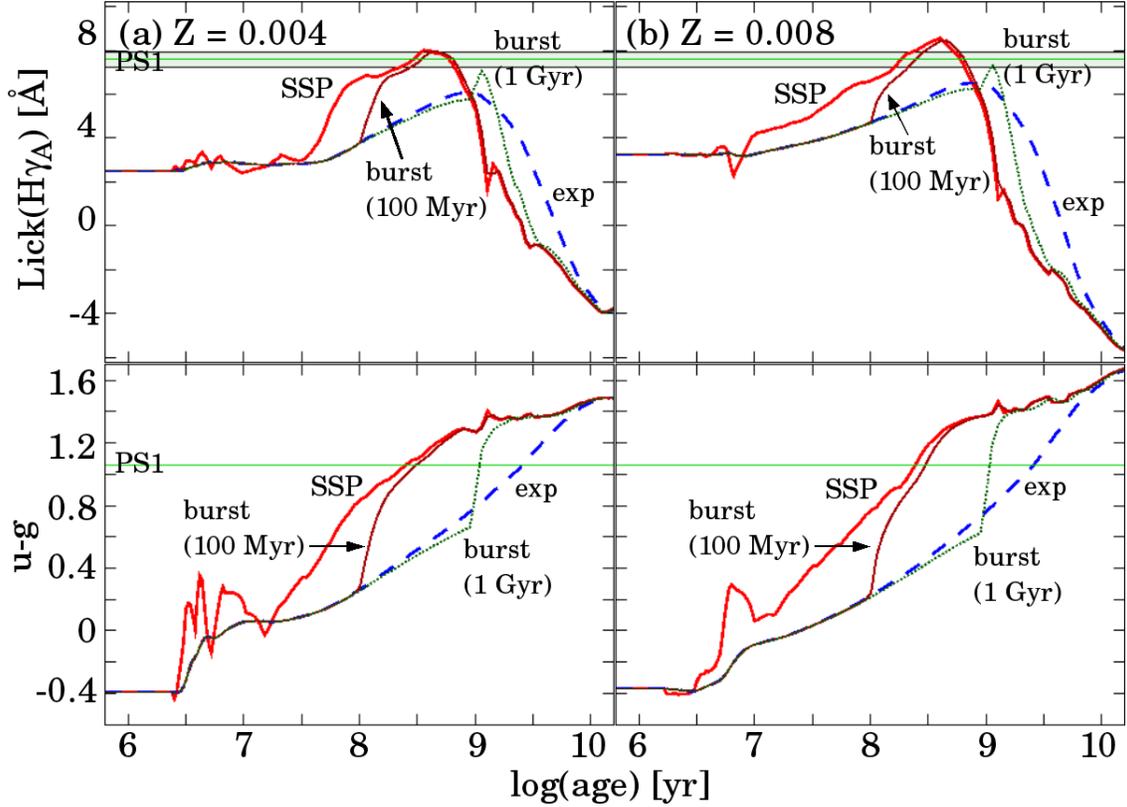}
\caption{H$\gamma_A$ Lick index ({\it top}) and u-g color index ({\it bottom})
predicted by four star-formation models \citep{Bruzual:2003} as a function of
starburst age, with the indices obtained in PS1.
The metallicities used in the models are (a) Z = 0.004 and (b) Z = 0.008,
respectively.
Bold solid, long-dashed, dotted, and faint solid lines represent the index of
the SSP, exp, burst (star-formation duration: 1 Gyr), and burst (star-formation
duration: 100 Myr) models, respectively (see text for detail).
The faint horizontal lines mark the observed H$\gamma_A$ Lick index by FOCAS
and u-g color in PS1.
The shaded regions in the H$\gamma_A$ index panels represent the values that
fall within 3 $\sigma$ for FOCAS data.
(A color version of this figure is available in the online journal.)
\label{fig:model-burst}}
\end{figure}

We carried out a least-squares fit for the two Lick indices and the four color
indices in PS1.
The best-fit model was 200 Myr in the SSP model for Z = 0.004 and 0.008.
The lowest $\chi^2$ values for the SSP model are one order of magnitude lower
than those of the other star-formation histories (Table \ref{tb:ew-color-fit}).
\citet{Yamauchi:2005} showed that the burst model, which has a constant
starburst with a duration of 1 Gyr at the beginning and no star formation
thereafter, is a plausible star formation history of E+A galaxies (their Fig.
13).
We compared the observed values also with the values predicted by this burst
model.
Figure \ref{fig:model-burst} shows the H$\gamma_A$ Lick index and u-g color
index predicted by SSP, exp, and burst models.
Burst model can produce larger H$\gamma_A$ index than exp model.
The u-g color of burst model at the age of peak H$\gamma_A$ index, $\sim$ 1.2
Gyr, is comparable to the observed u-g color.
However, even at that age, burst model cannot explain the large observed
H$\gamma_A$ index in $\sim$ 3 $\sigma$ level, due to the older stellar
population born before the end of star formation. 
Then, for reducing the contribution of the older stellar population, we changed
the star-formation duration of burst model into 100 Myr.
Predicted values of short burst model become almost the same as those of SSP
around 200 Myr after the burst begins, and this burst model at $\sim$ 300 Myr
fits the observed PS1 values (Table \ref{tb:ew-color-fit}).
Therefore, we conclude that PS1 is a post-starburst region and the star
formation in PS1 was suddenly quenched,
whether the best fit star-formation history is SSP or burst.

\begin{figure}
\epsscale{.80}
\plotone{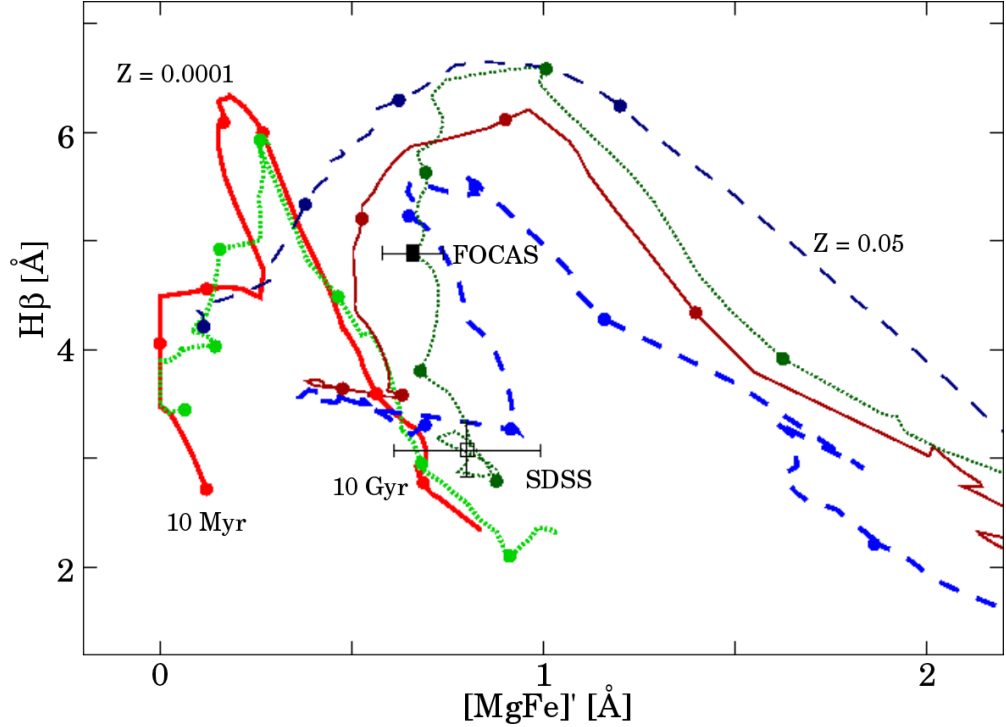}
\caption{The H$\beta$ Lick indices plotted against the [MgFe]\arcmin\, indices.
Open and filled squares represent the indices obtained for the central region
of J1602: by SDSS for the central 3\arcsec\, and by FOCAS for PS1, respectively,
with 1$\sigma$ error bars.
Lines represent the models \citep{Bruzual:2003}.
One line represents the evolutionary track for varying age at constant
metallicity.
The model metallicities are 0.0001 (bold solid), 0.0004 (bold dotted), 0.004
(bold long-dashed), 0.008 (faint solid), 0.02 (faint dotted), and 0.05 (faint
long-dashed), respectively.
Filled circles represent the ages at 10 Myr, 30 Myr, 100 Myr, 300 Myr, 1 Gyr,
3 Gyr, and 10 Gyr.
(A color version of this figure is available in the online journal.)
\label{fig:tmb-ps1}}
\end{figure}

We roughly estimated the metallicity in PS1 from the observed Lick indices of
the absorption lines.
In Fig. \ref{fig:tmb-ps1}, the H$\beta$ Lick index is plotted against the
[MgFe]\arcmin\, index, which is defined as
$\sqrt{{\rm Mg} b \times (0.72 \times {\rm Fe5270} + 0.28 \times {\rm
Fe5335})}$ \citep{Thomas:2003}.
These are good metallicity indicators, as the H$\beta$ index has little
sensitivity to element ratio variations or total metallicity, and the
[MgFe]\arcmin\, index is independent of $\alpha$/Fe \citep{Goto:2007}.
In this figure, we emphasized FOCAS data rather than SDSS data because the SDSS
spectrum contains a component of SB1 as well as PS1 due to its wide aperture
(3\arcsec).
Star-formation history used for the model is SSP.
Metallicity Z = 0.004, 0.008, or 0.02, which is comparable to the value
estimated from the [\ion{N}{2}]$\lambda$6583/H$\alpha$ flux ratio at SB1, can
explain observed index values.

\subsubsection{PS2}

The Lick indices and color indices of PS2 are close to those of PS1 except
for u-g color, which indicates that PS2 is also a post-starburst region.
Because of the bluer u-g color, the exp and const models at $\sim$ 1 Gyr can
also reproduce the observed Lick indices and color indices of PS2.
However, these models conflict with the fact that no emission line was found in
PS2.

\subsubsection{Two Population Model}
\label{sec:two-pop-model}

It is unlikely that J1602 has recently formed population only, because it would
be necessary to maintain large star-formation rate, e.g., more than 30
M$_{\odot}$ yr$^{-1}$ for 100 Myr, to produce observed stellar mass of J1602,
3.4 $\times$ 10$^9$ M$_{\odot}$.
Even larger star-formation rate would be required for the shorter
star-formation duration.
We roughly estimated the fraction of the stellar mass of post-starburst 
population to total stellar mass of J1602 using Lick index of H$\gamma_A$.
The observed index is almost the maximum value possible in the models, thus it
will put a severe constraint.
We assumed a combination of two stellar population for J1602: post-starburst
population (burst model) and old population (exp 10 Gyr).
The continuum flux was normalized with g-band magnitude, and metallicity
Z = 0.004 and 0.008 were used. 
For producing more than 6 \AA\, of H$\gamma_A$ index, more than $\sim$ 20 \% of
post-starburst population is needed, whether star formation duration of burst
model was 100 Myr or 300 Myr.
In order to produce this post-starburst population, star-formation rate of 11
M$_{\odot}$ yr$^{-1}$ or 4 M$_{\odot}$ yr$^{-1}$ is required
when masses of already dead early-type stars are also taken into consideration.
Although even 4 M$_{\odot}$ yr$^{-1}$ may be too large for J1602, whose stellar
mass is only 3.4 $\times$ 10$^9$ M$_{\odot}$, 300 Myr burst would be more
plausible.
Even larger H$\gamma_A$ index of 7.5 \AA\, is required at PS1, and more than
40 -- 70 \% of post-starburst population is needed.

\subsubsection{SB1 and SB2}

Many observed color indices in SB1 and SB2 are redder than those in PS1 and
PS2.
This is consistent with the SDSS image (Fig. \ref{fig:3dii-sdss}(a)):
the western side of the companion galaxy is redder than the eastern side.
We consider that dust formed by starbursts is reddening the western side.
The extinction and emission-line corrected indices in SB1 correspond to the exp
or const model at 100 Myr -- 1 Gyr (Fig. \ref{fig:color-model}).
These indices also correspond to the SSP model at 6 -- 100 Myr.
Taking further into consideration the observed H$\alpha$ and H$\beta$
equivalent widths (Table \ref{tb:FOCAS-reg2}) and a starburst model
\citep[Starburst99:][]{Leitherer:1999}, the const model at 300 -- 400 Myr or
the SSP model at $\sim$ 6 Myr is most likely.
The SB2 age is about the same as or slightly older than SB1 because of similar
color indices and smaller H$\alpha$ and H$\beta$ equivalent widths (Table 
\ref{tb:FOCAS-regA}).

SB1 probably has post-starburst population as well as young population.
If the post-starburst population greatly affects the observed color indices,
the conclusion with respect to SB1 star-formation history is invalid.
To estimate the effect of the post-starburst population, assuming that the
magnitudes of the post-starburst population in SB1 are the same as those in
PS2, which is the opposite side to SB1, we calculated that the fraction of flux
from the post-starburst population is only $\sim$ 30\% and color indices are
affected by only $\sim$ 0.07 mag.
Therefore, the conclusion of SB1 star-formation history derived from SB1 color
is valid.

\subsection{Evolutionary History of J1602}
\label{sec:discuss-history}

From our results, we suggest a history of J1602.
(A) Before galaxy interaction, both J1602 and its companion galaxy were
probably in only a moderate star-formation phase.
(B) They came close to each other and starbursts occurred within central 2 kpc
of J1602, including PS1, PS2, and maybe SB1. 
(C) Starbursts within central 2 kpc were suddenly quenched about 200 Myr ago
whereas starburst in SB1 may have continued.
(D) In the present state, PS1 and PS2 have become post-starburst regions
whereas SB1 is still or have become a starburst region.
(E) In the future, starburst at SB1 would eventually stop, and all of J1602
would become post-starburst regions.
If a gas mass of 10$^8$ M$_{\odot}$, for example, is continuously converted
into stars in the present star-formation rate in SB1 (0.25 M$_{\odot}$
yr$^{-1}$, Section \ref{sec:result-r2-focas}), the starburst will continue
further for about 400 Myr and J1602 will become a pure-E+A galaxy after about
600 Myr from now (i.e., about 200 Myr after the starburst stops).

We suggest two possible scenarios of starbursts in the second phase (B).
(B1) There was enough gas in all of J1602, and galaxy interaction
triggered starbursts within its central 2 kpc; or:
(B2) The gas was transported from the companion galaxy to J1602, and galaxy
interaction triggered starbursts within central 2 kpc.
This scenario, however, needs a large amount of gas transfer. 
Assuming that half of stars in PS1 were produced in phase (B) (Section 
\ref{sec:two-pop-model}), we calculated the produced stellar mass as 1.5
$\times$ 10$^8$ M$_{\odot}$ by using observed SDSS magnitude in PS1 and GALAXEV
\citep{Bruzual:2003}.
Even if the amount of gas in the companion galaxy is one-tenth of its stellar
mass, 1.5 $\times$ 10$^9$ M$_{\odot}$ (Section \ref{sec:obs-target}), the gas
mass will be equal to the required amount.
It would be difficult to transfer almost all gas in the companion galaxy since
the galaxy mass ratio of J1602 to the companion is only $\sim$ 2.
Therefore, we consider the (B1) scenario is more plausible.

We suggest three possible causes for starbursts to cease in the third phase (C).
(C1) All gas within the central 2 kpc was completely consumed, whereas gas in
SB1 may still have remained.
In this scenario, however, star formation would not fall sharply.
Thus, this scenario may not produce large Lick indices of H$\gamma_A$ and
H$\delta_A$.
(C2) Gas transport from the companion galaxy to J1602 ceased, and there was
no gas in J1602, whereas gas in SB1 may still have remained.
As mentioned above, however, gas transport from the companion galaxy is not
likely.
(C3) Galactic winds occurred and expelled gas from the central 2 kpc.
Galactic winds can quench a starburst suddenly and produce large Lick indices.
Although no signature of galactic winds was found in J1602 through our
observations, it would not be a serious problem because the time since the
galactic wind occurred in J1602 is much longer than the typical dynamical
timescales of galactic winds, which are shorter than 10 Myr (e.g.,
\citealt{Veilleux:2005,Matsubayashi:2009}).
Therefore, we consider the (C3) scenario is plausible.

\subsection{Further Details}
\label{sec:discuss-detail}

We should mention the timescales of galaxy rotation and galaxy movement
because these are comparable to the timescale between phase (C) and phase (D),
about 200 Myr.
Assuming that the dynamical mass of J1602 within central 2 kpc is equal to its
stellar mass, 3.4 $\times$ 10$^9$ M$_{\odot}$, we calculate that stars at PS2
make one rotation around J1602's center in 150 Myr.
Thus, the post-starburst population in PS2 was not necessarily born at the
apparent present place.
Taking into account the rotation of J1602, we suggest two possible
scenarios of the place where its starburst occurred in phase (B).
One scenario is that the starburst occurred in the whole central 2 kpc of
J1602.
This includes the case of an axisymmetric starburst distribution.
The other scenario is that the starburst occurred only at PS1 and the near side
to the companion galaxy \citep[e.g., NGC 6090:][]{Sugai:2004}.
Stars which had been born at the near side moved to the apparent present PS2
position by galaxy rotation and are now observed as a post-starburst population there.
We also roughly estimated timescale of the galaxy motion, although the
relative tangential velocity and the relative radial distance are uncertain.
It takes 30 Myr for galaxies to move the apparent distance between J1602 and
the companion galaxy, 4 kpc, with the observed relative radial velocity, 100 km
s$^{-1}$.
This means that J1602 and the companion in phase (C) were at a different
orbital position compared to phase (D).
Therefore, there is a possibility that the first pericenter triggered a
starburst in phase (B) and the second one triggered a starburst in phase (D).

J1602 and the companion galaxy are not severely disturbed in terms of
morphology (Fig. \ref{fig:image-focas}), unlike NGC 4038/9 (e.g.,
\citealt{Whitmore:1995}) or NGC 4676 (e.g., \citealt{Hibbard:1996}).
It may be because they did not experience a head-on collision and/or they are
at the beginning of a collision.
To guess whether the collision between these galaxies is prograde or
retrograde, we calculated the velocity in each region.
From Balmer absorption lines, the stellar velocity of PS2 relative to PS1
(i.e., the center of J1602) was calculated as +100 km s$^{-1}$.
Thus, J1602 rotates in a clockwise direction as viewed from north-west side if
we assume a simple galaxy rotation.
Since the velocity of SB2 (i.e., the center of the companion galaxy) relative
to PS1 was calculated as +200 km s$^{-1}$, we suggest that this collision
is retrograde.

Many galaxy-merger simulations (e.g., \citealt{Barnes:1996}) indicate that
gas falls into the center faster than stars.
Thus, it may be strange that in phase (D) SB1 is a starburst region but PS1
(the galactic center) is not.
\citet{Barnes:2004} indicates that shock-induced star formation provides a
better match to the observation of NGC 4676, an interacting galaxy system,
than the density-dependent model (e.g., \citealt{Mihos:1994}).
In the shock-induced models, more star formation occurs outside the center
compared with the density-dependent model.
There are some examples of off-center starbursts in interacting galaxy systems,
such as SDSS J101345.39+011613.66 \citep{Swinbank:2005}, SDSS
J161330.18+510335.5 \citep{Goto:2008}, and NGC 6090 \citep{Sugai:2004}.

Our observational data enable us to investigate the spatial distributions of
absorption and emission line regions, and to suggest a history of an
interesting galaxy, J1602.
However, we cannot completely understand J1602's history, e.g.,
the causes of starbursts in phase (B) and quenching of them in phase (C).
\ion{H}{1} or CO map observation of J1602 will be helpful.
Neutral/molecular gas emission will be detected mainly in SB1, not in PS1, if
scenarios (B1) and (C1) are true.
If a large amount of neutral/molecular gas emission is detected between J1602
and its companion galaxy, scenarios (B2) and (C2) would be plausible.
Follow-up observation of this interesting J1602 system, such as molecular
gas mapping, is important for examining the evolution of E+A galaxies.

\section{CONCLUSION}

In order to investigate the evolution of E+A galaxies, we observed SDSS J1602
and its neighbor galaxy with Kyoto3DII and FOCAS.
These are the first spatially resolved spectroscopic observations of E+A
progenitors.

We found a post-starburst region in the center of J1602 (PS1) and a starburst
region located 2 kpc from the center (SB1), in the direction of the neighbor
galaxy.
The fact that this galaxy has both starburst and post-starburst regions
indicates that it is in a critical phase of evolution.
The recession velocities of J1602 and its neighbor galaxy differ only by $\sim$
100 km s$^{-1}$.
Thus, they are a physical pair and are considered to have been interacting,
probably in a retrograde encounter.
A local velocity field of 80 km s$^{-1}$ is detected in SB1.

Comparing the observed Lick indices of the H$\gamma_A$ and H$\delta_A$ and
color indices to those predicted from stellar population synthesis models, we
find that a suddenly quenched star-formation scenario is plausible for the
star-formation history of PS1.
From our results, we suggest a history for J1602.
Starbursts occurred within 2 kpc from the center of J1602 probably due to
galaxy interaction, then about 200 Myr ago starbursts were suddenly quenched,
whereas the starburst in SB1 may have continued or have occurred recently.
The SB1 starburst will eventually stop, and all of J1602 will become a
post-starburst.
Follow-up observation, for example molecular gas mapping of this system, will
further elucidate the evolution process of E+A galaxies.

\acknowledgments

We appreciate the referee for many helpful comments.
We thank the staff at the UH 2.2-m telescope and the Subaru Telescope for their
help during the observation.
Use of the UH 2.2-m telescope for our observations was supported by National
Astronomical Observatory of Japan.
This work is based in part on data collected at Subaru Telescope, which is
operated by the National Astronomical Observatory of Japan.
We also thank J. E. Barnes for discussions and M. A. Malkan for valuable
comments and suggestions.
This work was partly supported by KAKENHI 21540247.

This work was supported by the Grant-in-Aid for the Global COE Program "The
Next Generation of Physics, Spun from Universality and Emergence" from the
Ministry of Education, Culture, Sports, Science and Technology (MEXT) of Japan.

Funding for the SDSS and SDSS-II has been provided by the Alfred P. Sloan
Foundation, the Participating Institutions, the National Science Foundation,
the U.S. Department of Energy, the National Aeronautics and Space
Administration, the Japanese Monbukagakusho, the Max Planck Society, and the
Higher Education Funding Council for England. The SDSS Web Site is
http://www.sdss.org/.

The SDSS is managed by the Astrophysical Research Consortium for the
Participating Institutions. The Participating Institutions are the American
Museum of Natural History, Astrophysical Institute Potsdam, University of
Basel, University of Cambridge, Case Western Reserve University, University of
Chicago, Drexel University, Fermilab, the Institute for Advanced Study, the
Japan Participation Group, Johns Hopkins University, the Joint Institute for
Nuclear Astrophysics, the Kavli Institute for Particle Astrophysics and
Cosmology, the Korean Scientist Group, the Chinese Academy of Sciences
(LAMOST), Los Alamos National Laboratory, the Max-Planck-Institute for
Astronomy (MPIA), the Max-Planck-Institute for Astrophysics (MPA), New Mexico
State University, Ohio State University, University of Pittsburgh, University
of Portsmouth, Princeton University, the United States Naval Observatory, and
the University of Washington. 

This research has made use of the NASA/IPAC Extragalactic Database (NED) which
is operated by the Jet Propulsion Laboratory, California Institute of
Technology, under contract with the National Aeronautics and Space
Administration.

This publication makes use of data products from the Two Micron All Sky Survey,
which is a joint project of the University of Massachusetts and the Infrared
Processing and Analysis Center/California Institute of Technology, funded by
the National Aeronautics and Space Administration and the National Science
Foundation.

\clearpage

\clearpage

\clearpage

\clearpage

\clearpage

\clearpage

\clearpage

\clearpage

\clearpage

\clearpage

\clearpage

\begin{deluxetable}{cc}
\tablecaption{Lick offsets applied for Kyoto3DII and FOCAS data
\label{tb:Lick-offsets}}
\tablewidth{0pt}
\tablehead{
\colhead{Index} & \colhead{Offset} \\
\colhead{} & \colhead{(\AA)} 
}
\startdata
H$\beta$ & 0.13 \\
Mg $b$ & 0.03 \\
Fe5270 & 0.17 \\
Fe5335 & 0.07 \\
H$\delta_A$ & 0.83 \\
H$\gamma_A$ & -0.89 \\
\enddata
\tablecomments{These values were subtracted from Lick indices calculated from
our data.}
\end{deluxetable}

\clearpage

\begin{deluxetable}{cccc}
\tablecaption{Observed values measured from the PS1 spectrum observed with
Kyoto3DII
\label{tb:3dii-reg1}}
\tablewidth{0pt}
\tablehead{
\colhead{Line} & \colhead{Wavelength} & \colhead{EW} & \colhead{FWHM} \\
\colhead{} & \colhead{(\AA)} & \colhead{(\AA)} & \colhead{(\AA)}
}
\startdata
H$\delta$ & 4278.6 & 6.1 & 13.7 \\
H$\gamma$ & 4526.8 & 7.1 & 16.9 \\
H$\beta$ & 5070.1 & 7.9 & 28.2 \\
\enddata
\end{deluxetable}

\clearpage

\begin{deluxetable}{ccccc}
\tablecaption{Observed values measured from the SB1 spectrum observed with
Kyoto3DII
\label{tb:3dii-reg2}}
\tablewidth{0pt}
\tablehead{
\colhead{Line} & \colhead{Wavelength} & \colhead{Flux} & \colhead{EW} &
\colhead{FWHM} \\
\colhead{} & \colhead{(\AA)} & \colhead{(10$^{-16}$ erg cm$^{-2}$ s$^{-1}$)} &
\colhead{(\AA)} & \colhead{(\AA)}
}
\startdata
H$\gamma$\tablenotemark{a} & 4527.2 & 4.4 & -6.4 & 3.6 \\
H$\beta$ & 5070.8 & 13.5 & -20.1 & 4.4 \\
{[\ion{O}{3}]}$\lambda$4959 & 5173.1 & 19.1 & -30.4 & 5.1 \\
{[\ion{O}{3}]}$\lambda$5007 & 5223.4 & 58.3 & -97.3 & 4.7 \\
\enddata
\tablenotetext{a}{This emission line was observed at a low S/N.}
\end{deluxetable}

\clearpage

\begin{deluxetable}{cccc}
\tablecaption{Lick indices in PS1 and PS2
\label{tb:FOCAS-Lick}}
\tablewidth{0pt}
\tablehead{
\colhead{Line} & \colhead{PS1 (Kyoto3DII)} & \colhead{PS1 (FOCAS)} &
 \colhead{PS2 (FOCAS)}
}
\startdata
H$\delta_A$ & --- & 6.78 $\pm$ 0.12 & 6.53 $\pm$ 0.36 \\
H$\gamma_A$ & 8.32 $\pm$ 0.61 & 7.58 $\pm$ 0.11 & 7.77 $\pm$ 0.36 \\
\enddata
\tablecomments{Uncertainties are in 1 $\sigma$ levels.
Lick offsets displayed in Table \ref{tb:Lick-offsets} were applied to the
observed values.
H$\delta_A$ index of PS1 observed with Kyoto3DII is not shown because a part of
the pseudo-continuum is out of the wavelength range in this observation.
}
\end{deluxetable}

\clearpage

\begin{deluxetable}{cccc}
\tablecaption{Observed values measured from the PS1 spectrum observed with
FOCAS
\label{tb:FOCAS-reg1}}
\tablewidth{0pt}
\tablehead{
\colhead{Line} & \colhead{Wavelength} & \colhead{EW} & \colhead{FWHM} \\
\colhead{} & \colhead{(\AA)} & \colhead{(\AA)} & \colhead{(\AA)}
}
\startdata
H13 & 3893.8 & 2.1 & 7.7 \\
H12 & 3909.6 & 2.6 & 7.8 \\
H11 & 3931.1 & 4.5 & 10.4 \\
H10 & 3959.7 & 6.4 & 13.5 \\
H9 & 3999.2 & 6.8 & 14.0 \\
H8 & 4055.2 & 8.0 & 16.7 \\
\ion{Ca}{2} K & 4100.7 & 1.4 & 11.3 \\
H$\varepsilon$ + \ion{Ca}{2} H & 4140.0 & 8.4 & 15.7 \\
H$\delta$ & 4277.6 & 6.6 & 14.5 \\
H$\gamma$ & 4526.7 & 7.6 & 18.3 \\
H$\beta$ & 5069.3 & 7.8 & 22.8 \\
\enddata
\end{deluxetable}

\clearpage

\begin{deluxetable}{ccccc}
\tablecaption{Observed values measured from the SB1 spectrum observed with
FOCAS
\label{tb:FOCAS-reg2}}
\tablewidth{0pt}
\tablehead{
\colhead{Line} & \colhead{Wavelength} & \colhead{Flux} & \colhead{EW} &
\colhead{FWHM} \\
\colhead{} & \colhead{(\AA)} & \colhead{(10$^{-16}$ erg cm$^{-2}$ s$^{-1}$)} &
\colhead{(\AA)} & \colhead{(\AA)}
}
\startdata
{[\ion{O}{2}]}$\lambda\lambda$3726,3729 & 3887.5 & 59.9 & -87.0 & 8.2 \\
H8 & 4057.2 & 2.7 & -4.1 & 7.1 \\
H$\varepsilon$ & 4140.9 & 2.0 & -2.8 & 6.0 \\
H$\delta$ & 4279.6 & 4.2 & -6.5 & 6.7 \\
H$\gamma$ & 4528.5 & 10.0 & -17.2 & 7.2 \\
{[\ion{O}{3}]}$\lambda$4363 & 4553.6 & 1.1 & -1.6 & 9.4 \\
H$\beta$ & 5071.4 & 24.5 & -47.3 & 7.0 \\
{[\ion{O}{3}]}$\lambda$4959 & 5173.3 & 29.6 & -51.8 & 6.9 \\
{[\ion{O}{3}]}$\lambda$5007 & 5223.2 & 88.7 & -156.5 & 6.8 \\
\ion{He}{1}$\lambda$5876 & 6129.6 & 3.6 & -7.7 & 6.8 \\
{[\ion{O}{1}]}$\lambda$6300 & 6572.6 & 1.7 & -4.0 & 7.1 \\
{[\ion{O}{1}]}$\lambda$6363 & 6639.1 & 0.6 & -1.4 & 6.4 \\
{[\ion{N}{2}]}$\lambda$6548 & 6831.1 & 2.9 & -7.2 & 6.8 \\
H$\alpha$ & 6846.9 & 90.9 & -247.9 & 6.8 \\
{[\ion{N}{2}]}$\lambda$6583 & 6867.8 & 7.6 & -22.7 & 7.2 \\
\ion{He}{1}$\lambda$6678 & 6967.4 & 0.9 & -2.5 & 6.5 \\
{[\ion{S}{2}]}$\lambda$6716 & 7006.8 & 7.9 & -22.5 & 7.4 \\
{[\ion{S}{2}]}$\lambda$6731 & 7021.9 & 5.7 & -15.8 & 7.0 \\
\enddata
\end{deluxetable}

\clearpage

\begin{deluxetable}{cccc}
\tablecaption{Observed values measured from the PS2 spectrum observed with
FOCAS
\label{tb:FOCAS-reg3}}
\tablewidth{0pt}
\tablehead{
\colhead{Line} & \colhead{Wavelength} & \colhead{EW} &
\colhead{FWHM} \\
\colhead{} & \colhead{(\AA)} & \colhead{(\AA)} & \colhead{(\AA)}
}
\startdata
H10 & 3960.8 & 4.1 & 10.1 \\
H9 & 3999.6 & 4.6 & 9.7 \\
H8 & 4056.0 & 3.4 & 11.1 \\
H$\varepsilon$ + \ion{Ca}{2} H & 4139.6 & 6.5 & 14.3 \\
H$\delta$ & 4279.7 & 6.5 & 17.3 \\
H$\gamma$ & 4527.5 & 7.2 & 19.2 \\
H$\beta$ & 5071.4 & 5.8 & 21.6 \\
\enddata
\end{deluxetable}

\clearpage

\begin{deluxetable}{ccccc}
\tablecaption{Observed values measured from the SB2 spectrum observed with
FOCAS
\label{tb:FOCAS-regA}}
\tablewidth{0pt}
\tablehead{
\colhead{Line} & \colhead{Wavelength} & \colhead{Flux} & \colhead{EW} &
\colhead{FWHM} \\
\colhead{} & \colhead{(\AA)} & \colhead{(10$^{-16}$ erg cm$^{-2}$ s$^{-1}$)} &
\colhead{(\AA)} & \colhead{(\AA)}
}
\startdata
{[\ion{O}{2}]}$\lambda\lambda$3726,3729 & 3887.5 & 7.7 & -33.5 & 10.0 \\
H$\gamma$ & 4529.6 & 0.7 & -3.1 & 7.0 \\
{[\ion{O}{3}]}$\lambda$4363 & 4550.8 & 0.8 & -3.2 & 9.1 \\
H$\beta$ & 5072.5 & 2.3 & -10.2 & 7.5 \\
{[\ion{O}{3}]}$\lambda$4959 & 5174.5 & 2.2 & -8.2 & 8.4 \\
{[\ion{O}{3}]}$\lambda$5007 & 5224.5 & 6.1 & -23.3 & 7.6 \\
\ion{He}{1}$\lambda$5876 & 6130.2 & 0.7 & -2.9 & 12.2 \\
{[\ion{O}{1}]}$\lambda$6300 & 6574.1 & 0.6 & -2.7 & 8.8 \\
{[\ion{N}{2}]}$\lambda$6548 & 6831.7 & 0.3 & -1.3 & 6.6 \\
H$\alpha$ & 6848.4 & 10.7 & -47.3 & 7.7 \\
{[\ion{N}{2}]}$\lambda$6583 & 6865.9 & 0.5 & -2.3 & 10.0 \\
{[\ion{S}{2}]}$\lambda$6716 & 7007.8 & 1.9 & -9.3 & 8.8 \\
{[\ion{S}{2}]}$\lambda$6731 & 7022.9 & 1.4 & -6.8 & 8.7 \\
\enddata
\end{deluxetable}

\clearpage

\begin{deluxetable}{ccccccc}
\tablecaption{Observed and corrected color indices of each region
\label{tb:color-sdss}}
\tablewidth{0pt}
\tablehead{
\colhead{Color index} & \colhead{PS1} & \colhead{SB1} & \colhead{SB1} &
\colhead{PS2} & \colhead{SB2} & \colhead{SB2}\\
\colhead{} & \colhead{(observed)} & \colhead{(observed)} &
\colhead{(corrected)\tablenotemark{a}} & \colhead{(observed)} &
\colhead{(observed)} & \colhead{(corrected)\tablenotemark{a}}
}
\startdata
u - g & 1.06 $\pm$ 0.01 & 0.92 $\pm$ 0.02 & 0.57 -- 0.71 & 0.75 $\pm$ 0.06 &
1.13 $\pm$ 0.05 & 0.52 -- 0.84 \\
g - r & 0.03 $\pm$ 0.00 & 0.02 $\pm$ 0.00 & -0.12 -- 0.01 & 0.15 $\pm$ 0.03 &
0.23 $\pm$ 0.02 & -0.27 -- 0.03 \\
r - i & 0.04 $\pm$ 0.01 & 0.22 $\pm$ 0.01 & -0.18 -- -0.09 & 0.02 $\pm$ 0.06 &
0.27 $\pm$ 0.03 & -0.14 -- 0.05 \\
i - z & 0.02 $\pm$ 0.03 & 0.04 $\pm$ 0.06 & -0.08 -- -0.01 & 0.18 $\pm$ 0.21 &
0.13 $\pm$ 0.09 & -0.14 -- 0.01 \\
\enddata
\tablecomments{Uncertainties are in 1 $\sigma$ levels.}
\tablenotetext{a}{The bluer indices represent the values when we assume
$E_{star}(B-V)$ = $E_{gas}(B-V)$ reddening ratio, whereas the redder indices
represent ones when we assume $E_{star}(B-V)$ = 0.44 $E_{gas}(B-V)$.
The u-g, g-r, and r-i color indices were corrected for the effects of internal
dust extinction and emission lines, whereas only the extinction correction was
made to the i-z color.}
\end{deluxetable}

\clearpage

\begin{deluxetable}{ccc}
\tablecaption{Best fit age and $\chi^2$ of each model
\label{tb:ew-color-fit}}
\tablewidth{0pt}
\tablehead{
\colhead{} & \colhead{log(age) [yr]} & \colhead{$\chi^2$}
}
\startdata
Z = 0.004 & & \\
SSP & 8.31 & 60.7 \\
exp & 9.01 & 1011 \\
const & 9.06 & 1507 \\
burst\tablenotemark{a} & 8.41 & 58.6 \\
Z = 0.008 & & \\
SSP & 8.31 & 155 \\
exp & 9.01 & 980 \\
const & 9.11 & 1519 \\
burst\tablenotemark{a} & 8.41 & 130 \\
\enddata
\tablenotetext{a}{The star-formation duration is 100 Myr.}
\end{deluxetable}

\clearpage

\end{document}